\documentclass[12pt]{iopart}
\usepackage{amssymb}

\begin{document}

\title[Quantum Partial Search of a Database with Several Target Items]
{Quantum Partial Search of a Database with Several Target Items}

\author{Byung-Soo Choi}
\address{Department of Electronics Engineering\\
School of Information and Communication Engineering \\
Sungkyunkwan University \\
Republic of Korea} \ead{bschoi3@gmail.com}
\author{Vladimir\ E.\ Korepin}
\address{C.N. Yang Institute for Theoretical Physics,\\
State University of New York,\\
Stony Brook, NY 11794-3840, USA}
\ead{korepin@max2.physics.sunysb.edu}

\begin{abstract}
We consider a database separated into blocks. Blocks containing
target items are called target blocks. Blocks without target items
are called non-target blocks. We consider a case, when each target
block has the same number of target items. We present a fast quantum
algorithm, which finds one of the target blocks. Our algorithm is
based on Grover-Radhakrishnan algorithm of partial search. We
minimize the number of queries to the oracle. We analyze the algorithm for blocks of 
large size  and also shows how to use the algorithm for finite blocks.
\end{abstract}

\pacs {03.67.-a, 03.67.Lx}

\submitto{\JPA}
\maketitle

\section{Introduction}

Database search has many applications. Search algorithm enters as a
subroutine in many important algorithms. Grover discovered a quantum
algorithm, which searchs faster than classical \cite{Grover}. It
starts searching from a uniform superposition of all input states in
the database (\ref{ave}). Grover algorithm was proven to be optimal
\cite{Bennett, Boyer}. If a database has one target item
Grover algorithm can find it in
\begin{equation}
j_{\mbox{full}}= \frac{\pi}{4} \sqrt{N}, \qquad N\rightarrow \infty \label{full}
\end{equation}
queries  [iterations]. If the database has $z$ target items and we
want to find one of them, Grover algorithm can find it in
\begin{equation}
j_{\mbox{full}}= \frac{\pi}{4} \sqrt{\frac{N}{z}}, \qquad N\rightarrow \infty \label{mult}
\end{equation}
queries to the oracle, see \cite{Boyer}. This is a full search.

Partial search considers the following problem: a database is
separated into $K$ {blocks}, of a size
\begin{equation}
b=\frac{N}{K} \label{size}.
\end{equation}
A user wants to find one target block, which contains the target
item. Grover and Radhakrishnan discovered a quantum algorithm for
partial search \cite{jaik}. A general structure of the algorithm is
\begin{itemize}
\item Step 1. Global queries: $j_1$ standard Grover iterations, see (\ref{iter}).
\item Step 2. Simultaneous local searches in each block: $j_2$ local Grover iterations, see (\ref{liter}).
\item Step 3. One reflection over the average amplitude in the whole database, see (\ref{average}).
\end{itemize}
Partial search also starts from a uniform superposition of all basis
states (\ref{ave}). For many large blocks an improved version of
partial search algorithm was suggested in \cite{kg}. The number of
queries to the oracle for several blocks of large size was minimized
in \cite{kor}. This is GRK algorithm which we shall use in this
paper. Other partial search algorithms were considered in \cite{kl}.
 GRK algorithm was formulated in terms of group theory in \cite{val}, this provides a powerful mathematical tool for proof of optimality.
Partial search was first discovered while study a  list matching, see \cite{hei}.
The GRK algorithm was formulated  for blocks of finite size in
\cite{quant-ph-0603136}. Let us emphasize that only one target block
with one target item was considered in
\cite{jaik,kg,kor,kl,quant-ph-0603136}. To see the universal
features of the algorithm we consider large blocks $b\rightarrow
\infty$. As the size of the blocks increases the numbers of queries
during each Step $j_1$ and $j_2$ are controlled by parameters $\eta$
and $\alpha$, see (\ref{steps}). We choose these parameters to
minimize the total number of queries $j_1+j_2$, see (\ref{answer}).

In this paper we generalize GRK algorithm to the following. The
database has $N$ items, it is separated into $K$ blocks. Each block
has $b$ items, so $N=Kb$. Some of the blocks are target blocks, and
we denote the number of target blocks by $t$. Each target block has
several target items, and we denote the number of target items in a
target block by $\tau$ (same for each target block). Total number of target items in the whole
database is $z=t\tau$. We shall denote a target item by $|m\rangle$.
A collection of all target items in the database is
$A=\{|m\rangle\}$, a collection of all target items in a target
block is $a=\{|m\rangle\}$. Non-target items are $|x\rangle$, a
collection of all non-target items in the database is
$X=\{|x\rangle\}$, a collection of all non-target items in a block
is ${\rm x}=\{|x\rangle\}$. So the whole database is $D=A\cup X$.
Non-target block $B={\rm x}$, but target block $B={\rm x}\cup a$. We
optimize Grover-Radhakrishnan algorithm to find one of the target
blocks.

This paper consists of two parts. In the first part we consider the
limit of large blocks $b\rightarrow \infty$. We optimize Grover
Radhakrishnan algorithm to find one of the target blocks. We compare
our result with GRK algorithm \cite{kor} and show that the optimal
version of partial search in a database with several target blocks
can be obtained from GRK [one target block with one target item] by
replacing $b$ by $b/\tau$ and $K$ by $K/t$ , see
(\ref{re1}-\ref{re2}). In the second part we consider 
application of partial search to  finite blocks. For this case, we
find a sure success way.

\section{The Partial Search Algorithm}

\subsection{Step 1: Global Queries}

The Grover algorithm is based on two operations. First one
\begin{equation}
I_t=\hat{I}-2 \sum_{m\in A}^{t\tau} |m\rangle \langle m| \label{target}
\end{equation}
selectively inverts the amplitude of the states of target items
$|m\rangle$, where $\hat{I}$ is the identical operator. Initial
state of the database is a uniform superposition of all basis states
as
\begin{equation}
|s_1\rangle = \frac{1}{\sqrt{N}}\sum_{x=0}^{N-1}|x\rangle , \qquad
\langle s_1|s_1\rangle =1. \label{ave}
\end{equation}
Here $N$ is the total number of items in the whole database. Second
operation
\begin{equation}
I_{s_1}=\hat{I}-2|s_1\rangle \langle s_1| \label{average}
\end{equation}
selectively inverts the amplitude of the uniform superposition.
Consider a vector
\begin{equation}
|v\rangle=\sum_{x=0}^{N-1}a_x|x\rangle.
\end{equation}
The operator $-I_{s_1}$ inverts the coefficients about the average
as
\begin{eqnarray}
-I_{s_1}|v\rangle=\sum_{x=0}^{N-1}\mbox{\u{a}}_x|x\rangle, \quad
\mbox{\u{a}}_x =2\bar{a}-a_x, \quad
\bar{a}=\frac{1}{N}\sum_{x=0}^{N-1} {a_x}. \label{reflect}
\end{eqnarray}
Finally, Grover iteration is a unitary operator as
\begin{equation}
G_1=-I_{s_1}I_t. \label{iter}
\end{equation}
From now, we shall call it a global iteration. We have to apply it
several times $G_1^{j_1}$. To describe global query more precisely
we shall need an angle $\theta_1$ defined by
\begin{equation}
\sin ^2\theta_1 =\frac{t\tau}{N}. \label{ang1}
\end{equation}
We assume that initially our database is in uniform superposition
(\ref{ave}) as $|s_1\rangle \label{nach}$

During the first step we use only standard Grover iterations
(\ref{iter}). We shall use eigenvectors of $G_1$ as
\begin{eqnarray}
 G_1|\psi^{\pm}_1\rangle & =& \lambda^{\pm}_1 |\psi^{\pm}_1\rangle , \qquad
\lambda^{\pm}_1 =\exp[{\pm 2i \theta_1}] , \label{value}  \\
 |\psi^{\pm}_1\rangle &  =& \frac{1}{\sqrt{2}}\left(\frac{1}{\sqrt{t\tau}}\sum^{t\tau}_{m\in A} |m\rangle\right) \pm \frac{i}
{\sqrt{2}}\left( \frac{1}{\sqrt{(N-t\tau)}}\sum^{N-t\tau}_{x\in
X}|x\rangle \right). \nonumber
\end{eqnarray}
They were found  in
\cite{2000quant.ph..5055B}. The first Step of the partial search
consists of $j_1$ standard Grover iterations applied to the whole
database. After $j_1$ queries the wave-function of the database is
\begin{eqnarray}
G_1^{j_1} |s_1\rangle  &=\frac{ \sin \left( (2j_1+1)\theta_1
\right)}{\sqrt{t\tau}} \sum_{x\in A}^{t\tau}|m\rangle& + \frac{\cos
\left( (2j_1+1)\theta_1 \right)}{\sqrt{N-t\tau}}
\sum^{N-t\tau}_{x\in X}|x\rangle.\label{first}
\end{eqnarray}
It also can be found in \cite{2000quant.ph..5055B}.

\subsection{Step 2: Local Queries}

During the second Step local searches are made in each block separately in parallel.
We can write it as a direct sum with respect to all blocks
\begin{equation}
\hat{G_2}=\oplus G_2.
\end{equation}
In each block local search is a Grover iteration as
\begin{equation}
G_2=-I_{s_2}I_t .\label{liter}
\end{equation}
Here
\begin{equation}
I_t =\hat{I}-2\sum_{m\in a}^\tau |m\rangle\langle m|
\end{equation}
and  $I_{s_2}$ is
\begin{equation}
I_{s_2}=\hat{I}-2|s_2\rangle \langle s_2|, \qquad |s_2\rangle =
\frac{1}{\sqrt{b}}\sum_{\mbox{\scriptsize{one block}}}|x\rangle,
\qquad \langle s_2|s_2\rangle=1. \label{nblock}\end{equation}
Physical implementation of local search is similar to implementation the standard Grover search.
Local search is the Grover search in a block.
Remember that the number of items in each block is $b=N/K$. First
let us consider the wave-function of a non-target block  [a projection of the wave function of the database to this block] as
\begin{equation}
|nB\rangle =a_{nt} \sum_{\mbox{\scriptsize{one block}} }|x\rangle =
a_{nt}\  \sqrt{b} \ |s_2\rangle.
\end{equation}
We can read the coefficient from (\ref{first}) as
\begin{equation}
a_{nt} = \frac{\cos \left( (2j_1+1)\theta_1
\right)}{\sqrt{N-t\tau}}. \label{nt}
\end{equation}
The wave-function of a non-target block does not change during the
second step as
\begin{equation}
I_t|nB\rangle =|nB\rangle =-I_{s_2}|nB\rangle =G_2 |nB\rangle .
\end{equation}
Now let us consider the target block. After Step 1 its wave function
can be obtained from (\ref{first}) as
$$|B\rangle=\frac{\sin \left( (2j_1+1)\theta_1 \right)}{\sqrt{t}}  |\mu\rangle +
{}\frac{\cos \left( (2j_1+1)\theta_1
\right)\sqrt{b-\tau}}{\sqrt{N-t\tau}} |ntt\rangle. $$
Here $|\mu\rangle$ is the normalized sum of target items as
\begin{equation}
|\mu\rangle={\frac{1}{\sqrt{\tau}}}\sum_{m\in a}^\tau
|m\rangle,\qquad \langle \mu|\mu \rangle=1
\end{equation}
and $|ntt\rangle $ is a normalized sum of all non-target items in a
target block as
\begin{equation}
|ntt\rangle = \frac{1}{\sqrt{b-\tau}} \sum^{b-\tau}_{\stackrel{x \in
{\rm x}}{\mbox{\tiny{target block}}}} |x\rangle,\qquad \langle ntt
|ntt\rangle =1 .\label{vntt}
\end{equation}
Note that we can express the state (\ref{nblock}) as
\begin{equation}
|s_2\rangle = \sqrt{\frac{\tau}{b}} \quad |\mu\rangle +
\sqrt{\frac{b-\tau}{b}} \quad |ntt \rangle.
\end{equation}
We shall need an angle $\theta_2$ as
\begin{equation}
\sin^2 \theta_2 =\frac{\tau}{b}. \label{ang2}
\end{equation}
Eigenvectors of $G_2$ are constructed similar to eigenvectors of
$G_1$, see (\ref{value}), as
\begin{eqnarray}
 G_2|\psi^{\pm}_2\rangle &=&\lambda^{\pm}_2 |\psi^{\pm}_2\rangle, \qquad
 \lambda^{\pm}_2  = \exp[{\pm 2i \theta_2}] , \label{vector2}  \nonumber \\
|\psi^{\pm}_2\rangle  & =&\frac{1}{\sqrt{2}}|\mu\rangle \pm
\frac{i}{\sqrt{2}} |ntt\rangle.
\end{eqnarray}
We can resolve these equations as
$$|ntt\rangle =  \frac{i}{\sqrt2} \left(- |\psi^{+}_2\rangle + |\psi^{-}_2\rangle \right), \quad
 |\mu\rangle = \frac{1}{\sqrt2}\left( |\psi^{+}_2\rangle + |\psi^{-}_2\rangle
 \right).$$
Now we can express the wave function of the target block in terms of
the eigenfunctions of $G_2$ as
\begin{eqnarray}
|B\rangle = &\left[\frac{\sin \left( (2j_1+1)\theta_1
\right)}{\sqrt{ 2t}} -
i\sqrt{\frac{b-\tau}{N-t\tau}} \frac{\cos \left( (2j_1+1)\theta_1 \right)}{\sqrt 2}\right] |\psi^{+}_2\rangle + \nonumber \\
&\left[\frac{ \sin \left( (2j_1+1)\theta_1 \right)}{\sqrt{2t}} +
i\sqrt{\frac{b-\tau}{N-t\tau}} \frac{\cos \left( (2j_1+1)\theta_1
\right)}{\sqrt 2} \right]|\psi^{-}_2\rangle.
\end{eqnarray}
After $j_2$ iterations the target block will be
\begin{eqnarray}
&G^{j_2}_2|B\rangle = \nonumber \\
&\frac{e^{2i\theta_2j_2}}{\sqrt 2} \left[{\frac{\sin \left(
(2j_1+1)\theta_1 \right) }{\sqrt{t}}}-
i\sqrt{\frac{b-\tau}{N-t\tau}} {\cos \left( (2j_1+1)\theta_1 \right)}\right] |\psi^{+}_2\rangle+  \nonumber \\
&\frac{e^{-2i\theta_2j_2}}{\sqrt2} \left[{\frac{ \sin \left(
(2j_1+1)\theta_1 \right)}{\sqrt{t}} }+i\sqrt{\frac{b-\tau}{N-t\tau}}
{\cos \left( (2j_1+1)\theta_1 \right)} \right]|\psi^{-}_2\rangle.
\end{eqnarray}
Now we can use (\ref{vector2}) to express the wave functions of the
target block in the following form
\begin{equation}
|B_2\rangle \equiv  G^{j_2}_2|B\rangle = a_t |\mu\rangle
+a_{ntt}|ntt\rangle. \label{stage}
\end{equation}
and to calculate the amplitudes
\begin{eqnarray}
a_t =&\frac{\cos (2j_2\theta_2) \sin \left( (2j_1+1)\theta_1 \right)}{\sqrt{t}} + \nonumber \\
&\sqrt{\frac{b-\tau}{N-t\tau}}
\sin (2j_2\theta_2)\cos \left( (2j_1+1)\theta_1 \right) , \label{at} \\
a_{ntt}= &-\frac{\sin (2j_2\theta_2) \sin \left( (2j_1+1)\theta_1
\right)}{\sqrt{t}} +
\nonumber \\
&\sqrt{\frac{b-\tau}{N-t\tau}} \cos (2j_2\theta_2)\cos \left(
(2j_1+1)\theta_1 \right) . \label{ntt}
\end{eqnarray}
Later we shall also use
\begin{equation}
I_t|B_2\rangle =- a_t |\mu\rangle +a_{ntt}|ntt\rangle \label{cel}
\end{equation}
\subsection{Step 3: Location of a Target Block }

The last Step consists of the following operation. We start by the
application of the operator
\begin{equation}
-I_{s_1}=-\hat{I}+2|s_1\rangle \langle s_1|.
\end{equation}
[see (\ref{ave})] to the wave function of the whole database. This
is a global inversion about the average ${a}_x \rightarrow
2\bar{a}-a_x$, see (\ref{reflect}).

The amplitude of an item which we want to vanish should be double of
the average: $a_x=2\bar{a}$. We want to annihilate the amplitudes of
items in the non-target blocks $a_{nt}\rightarrow 0$. This means
that the amplitudes introduced in the previous subsection should
satisfy $a_{nt}=2\bar{a}$ as
\begin{equation}
a_{nt}={\frac{2}{N}} \left[ b ( K-t
)a_{nt}+t\sqrt{\tau}a_t+t\sqrt{b-\tau} a_{ntt} \right]
\label{annihilation}
\end{equation}
see (\ref{vntt}). We can rewrite this as
\begin{equation}
b \left( -\frac{K}{2} +t
\right)a_{nt}=t\sqrt{\tau}a_t+t\sqrt{b-\tau}a_{ntt}.
\label{cancelation}
\end{equation}
Substituting here the expressions (\ref{nt}), (\ref{at}),
(\ref{ntt}) we obtain the cancellation equation as
\begin{eqnarray}
&\frac{b\left(-K/2+t \right)}{\sqrt{N-t\tau}} \cos \left( (2j_1+1)\theta_1 \right) = \nonumber \\
&\sqrt{t\tau}\cos (2j_2\theta_2)\sin \left( (2j_1+1)\theta_1 \right)+ \nonumber \\
&t\sqrt{\tau\frac{b-\tau}{N-t\tau}} \sin (2j_2\theta_2)\cos \left( (2j_1+1)\theta_1 \right)-  \nonumber  \\
&\sqrt{t(b-\tau)}\sin (2j_2\theta_2) \sin \left( (2j_1+1)\theta_1 \right)+ \nonumber \\
&t{\frac{b-\tau}{\sqrt{N-t\tau}}} \cos (2j_2\theta_2)\cos \left(
(2j_1+1)\theta_1 \right). \label{center}
\end{eqnarray}
This equation guarantees that the amplitude of each item in each
non-target block vanishes as
\begin{eqnarray}
 &-I_{s_1}|nB\rangle=0, \label{uspeh} \\
& -I_{s_1}|B_2\rangle = -(a_t-\sqrt{\tau}a_{nt})|\mu
\rangle -(a_{ntt}-\sqrt{b-\tau} a_{nt})|ntt\rangle. \nonumber
\end{eqnarray}
Here we used (\ref{stage}), (\ref{reflect}) and $a_{nt}=2\bar{a}$.

Now we can do a {\it measurement}. In the simplest case $N=2^n$ and
$K=2^k$, we label blocks by $k$ qubits[items inside of a block are
labeled by $n-k$ qubits]. We measure only $k$ block qubits and {\bf
find a target block}.

Let us make an extra query to the oracle to put a target block into
a canonical form as
\begin{eqnarray}
 |B_3\rangle&=-I_tI_{s_1}|B_2\rangle = \sin \omega |\mu\rangle +\cos  \omega |ntt\rangle \label{fin} \\
   & =(a_t-\sqrt{\tau}a_{nt})|\mu\rangle -(a_{ntt}-\sqrt{b-\tau}
a_{nt})|ntt\rangle. \label{konez}
\end{eqnarray}
In later, we shall calculate the angle $\omega$, see (\ref{ugol}).

Meanwhile, we can also apply last two operations in inverse order as
follows.
\begin{equation}
\quad |{\cal B}_3\rangle =-I_{s_1}I_t|B_2\rangle =
(a_t+\sqrt{\tau}a_{nt})|\mu\rangle -(a_{ntt}-\sqrt{b-\tau}
a_{nt})|ntt\rangle \label{choi}
\end{equation}
Now  to vanish amplitudes of items in non-target blocks
(\ref{uspeh}) we will have to change the sign of $a_t$ in equations
(\ref{annihilation}) and (\ref{cancelation}), compare (\ref{stage})
and (\ref{cel}). Also the sign of first two terms in the right hand
side of eq (\ref{center}).

\subsection{The Limit of Large Blocks}

To see universal features we consider the limit when each block is
very large $b\rightarrow \infty$, this makes the total number of
items in the whole database also large
 $ N=Kb\rightarrow \infty$.
The expressions for angles (\ref{ang1}), (\ref{ang2}) simplifies as
\begin{equation}
\theta_1=\sqrt{\frac{t\tau}{{N}}},\qquad \theta_2=\sqrt{\frac{\tau}{{b}}}.\label{dva}
\end{equation}
The numbers of global and local iterations scales as, respectively
\begin{equation}
j_1=\frac{\pi}{4}\sqrt{\frac{N}{t\tau}} -\eta \sqrt{b} ,\qquad
j_2=\alpha \sqrt{b}, \label{steps}
\end{equation}
similar to $t=1$, $\tau=1 $ case \cite{jaik,kor,kl}. Here $\eta$ and
$\alpha$ are parameters of order of 1 [they have a limit as
$b\rightarrow \infty$]. We shall minimize the total number of
queries and find the best values of these parameters as a functions
of number of blocks $K$, see (\ref{answer}). This will give the
optimal distribution of number of queries between Steps 1 and 2 of
the partial search algorithm. We shall need the arguments of
trigonometric functions in (\ref{center}) as
\begin{equation}
(2j_1+1)\theta_1=\frac{\pi}{2}- {2\eta\sqrt{\frac{t\tau}{{K}}}} ,\qquad  2j_2\theta_2=2\alpha\sqrt{\tau} . \label{arguments}
\end{equation}
Let us leave only leading terms [of order $\sqrt{b}$] in the
equation (\ref{center}) as $b\rightarrow \infty$
\begin{eqnarray}
&{\sqrt{b}} \left(- \frac{\sqrt{K}}{2} +\frac{t}{\sqrt{K}} \right)\sin \left(
\frac{2\eta\sqrt{t\tau}}{\sqrt{K}} \right) \nonumber  \\
& = -\sqrt{bt} \sin (2\alpha \sqrt{\tau}) \cos \left( \frac{2\eta
\sqrt{t\tau}}{\sqrt{K}} \right) + t{\sqrt{\frac{{b}}{K}}} \cos
(2\alpha\sqrt{\tau})\sin \left(
\frac{2\eta\sqrt{t\tau}}{\sqrt{K}}\right) . \nonumber
\end{eqnarray}
Note that $a_t$ does not contribute in this limit, so this equation
is the same for both cases of preparation of the final state of a
target block (\ref{fin}) and (\ref{choi}).

We can simplify the equation to
\begin{eqnarray}
\left( \frac{\sqrt{K}}{2} -\frac{t}{\sqrt{K}} +\frac{t\cos(2\alpha
\sqrt{\tau})}{\sqrt{K}}\right)\sin \left(
\frac{2\eta\sqrt{t\tau}}{\sqrt{K}} \right) \\
= \sqrt{t}\sin (2\alpha\sqrt{\tau}) \cos \left(
\frac{2\eta\sqrt{t\tau}}{\sqrt{K}} \right). \nonumber
\end{eqnarray}
We can express $\eta$ as a function of $\alpha$ as
\begin{equation}
\tan \left(\frac{2\eta\sqrt{t\tau}}{\sqrt{K}} \right) =\frac{2\sqrt{Kt}\sin (2\alpha\sqrt{\tau)}}{K
 -4t\sin^2 (\alpha \sqrt{\tau)}} , \label{expr}
\end{equation}
see (\ref{steps}). Notice that after we introduce rescaled variables
\begin{equation}
\tilde{K}=\frac{K}{t}, \quad \tilde{\alpha}=\alpha\sqrt{\tau},\quad
\tilde{\eta}=\eta \sqrt{\tau}, \label{scale}
\end{equation}
the equation (\ref{expr}) will coincide with corresponding equation
from \cite{kor}
\begin{equation}
\tan \left(\frac{2\tilde{\eta}}{\sqrt{\tilde{K}}} \right) =\frac{2\sqrt{\tilde{K}}\sin (2\tilde{\alpha})}{\tilde{K}
 -4\sin^2 (\tilde{\alpha} )}. \label{k}
\end{equation}

\subsection{Optimization}

The most efficient algorithm makes the least number of queries to
the oracle as
\begin{eqnarray}
j_1+j_2+1\rightarrow \frac{\pi}{4}\sqrt{\frac{N}{t\tau}} +(\alpha
-\eta)  \sqrt{b} =\frac{\pi}{4}\sqrt{\frac{N}{t\tau}}
+\sqrt{\frac{b}{\tau}} (\tilde{\alpha} -\tilde{\eta})
\label{number}.
\end{eqnarray}
To optimize the algorithm we have to minimize
$$\tilde{\alpha} -\tilde{\eta},$$ with constraint (\ref{k}).
The optimal values of $\tilde{\alpha} $ and $\tilde{\eta}$ can be
copied from \cite{kor}. The minimum number of queries can be
achieved when
\begin{equation}
\tan \frac{2\tilde{\eta}(\tilde K)}{\sqrt{\tilde K}}=\frac{\sqrt{3\tilde{K}-4}}{\tilde{K}-2}, \qquad
\cos 2\tilde{\alpha} (\tilde{K})=\frac{\tilde{K}-2}{2(\tilde{K}-1)} . \label{answer}
\end{equation}
{This describes the optimal distribution of queries between Step 1
and Step 2}, see (\ref{steps}). We shall consider these formulae as
an expression of $\tilde{\eta}$ and $\tilde{\alpha}$ as functions of
$\tilde{K}$.

Let us summarize the result. The optimal number of global queries at
Step 1 is
\begin{equation}
j_1= \left\{  \frac{\pi}{4}\sqrt{\frac{K}{t}}- {\tilde{\eta}} \left( \frac{K}{t} \right) \right\}{\sqrt{\frac{b}{\tau}}}  .\label{re1}
\end{equation}
Here the function
\begin{equation}
\tilde{\eta}\left( \frac{K}{t}\right)=\frac{1}{2} \sqrt{\frac{K}{t} }\arctan  \left( \frac{\sqrt{t(3K-4t)}}{K-2t} \right) \label{f1}
\end{equation}
depends only on the ratio $K/t$. The number of local queries at Step
2 is
\begin{eqnarray}
j_2 &=& \tilde{\alpha} \left( \frac{K}{t}\right)
\sqrt{\frac{b}{\tau}},  \label{re2}
\end{eqnarray}
where
\begin{eqnarray}
\tilde{\alpha} \left( \frac{K}{t}\right) & = & \frac{1}{2} \arccos
\left(\frac{K-2t}{2(K-t)} \right).    \label{f2}
\end{eqnarray}

After Step 3 amplitudes of all items in all non-target blocks
vanish.

Let us discuss the final wave function of the target block
(\ref{fin}). Note that $a_{nt}$ does not contribute in the
coefficient in front of $|\mu \rangle$ as $b\rightarrow \infty$
 see (\ref{nt}), so wave functions (\ref{fin}) and (\ref{choi})
 coincide as
\begin{eqnarray}
& |{\cal B}_3\rangle=|B_3\rangle= \sin \omega |\mu \rangle +\cos
\omega |ntt \rangle. \label{nfin}
\end{eqnarray}
At the minimum (\ref{answer}) we get
\begin{equation} \omega=
\tilde{\alpha} \left( \frac{K}{t}\right).\label{ugol}
\end{equation}

\section{Sure success partial search}

\subsection{Motivation and Idea}

Until now, we have focused on how to find a target block in limit
of large blocks. The algorithm should be modified for  
 case of finite blocks.. Hence we should find a way for a realistic case when
$b$ is finite. In this section, we explain a sure success way for
this case.

Briefly, we will not change the total number of queries to the
oracle $j_1+j_2+1$, we will only change Step 3. In Step 3, we shall
modify both operators $I_{s_1}$ and $I_t$ by phases in such a way
that the algorithm will find a target block with $100 \%$
probability.

\subsection{Sure success way}

A sure success way for the case of single target item/single target
block partial search was already proposed in
\cite{quant-ph-0603136}. Therefore we can extend this sure success
way for our problem, partial search with multiple target
items/multiple target blocks. As in \cite{kl,quant-ph-0603136}, we
note that the entire action of the partial search can be compactly
described by a 3-dimensional subspace spanned by the vectors:
$|M\rangle$ the normalized state of all target items; $|NTT\rangle$
the normalized state of all non-target items in all target blocks;
and $|u\rangle$ the normalized state for all other items. Three
bases are respectively
\begin{eqnarray}
|M\rangle&=\frac{1}{\sqrt{t}}\sum_{\mbox{target
blocks}}^t|\mu\rangle=\frac{1}{\sqrt{t\tau}}\sum^{t\tau}_{m\in A} |m\rangle , \\
|NTT\rangle&=\frac{1}{\sqrt{t}}\sum_{\mbox{target
blocks}}^t|ntt\rangle=
\frac{1}{\sqrt{t(b-\tau)}}\sum_{\mbox{target
blocks}}^t\sum^{b-\tau}_{x \in
{\rm x}} |x\rangle, \\
|u\rangle&=\frac{1}{\sqrt{b(K-t)}}\sum_{\mbox{items in non-target
blocks}}^{b(K-t)}|x\rangle.
\end{eqnarray}
Using these orthonormal basis, the initial state $|s_1\rangle$ of
~(\ref{ave}) may be written
\begin{eqnarray}
|s_1\rangle&=&\sin\gamma\sin\theta_2|M\rangle+\sin\gamma
\cos\theta_2|NTT\rangle+\cos\gamma|u\rangle\;,
\end{eqnarray}
where $\sin^2\gamma=t/K\;$, $\sin^2\theta_1=t\tau/N$, and
$\sin^2\theta_2=\tau/b$ . Note that these three bases states are
generalized from \cite{quant-ph-0603136}. From this, all
subsequent flows are the same with the result in
\cite{quant-ph-0603136}.

Based on the three bases, $G^{j_1}_1$ operator is represented as
\begin{equation}
G^{j_1}_1= TM_{j_1}T,
\end{equation}
where
\begin{equation}
T= \left(
\begin{array}{ccc}
1&0&0\\
0&\frac{\cos\theta_2 \sin\gamma}{\cos\theta_1}&\frac{\cos\gamma}{\cos\theta_1} \\
0&\frac{\cos\gamma}{\cos\theta_1}&-\frac{\cos\theta_2
\sin\gamma}{\cos\theta_1}
\end{array} \right)
\end{equation}
and
\begin{equation}
M_{j_1}= \left(
\begin{array}{ccc}
\cos(2j_1\theta_1)&\sin(2j_1\theta_1)&0\\
-\sin(2j_1\theta_1)&\cos(2j_1\theta_1)&0\\
0&0&(-1)^{j_1}
\end{array} \right)\;.
\end{equation}
Also, based on the same three bases, $G^{j_2}_2$ operator is
represented as
\begin{equation}
G^{j_2}_2=\left(
\begin{array}{ccc}
\cos(2j_2\theta_2)  &   \sin(2j_2\theta_2)  &   0 \\
-\sin(2j_2\theta_2) &   \cos (2j_2\theta_2) &   0 \\
0                   &                   0   &   1
\end{array}
\right)\;.
\end{equation}

The intermediate state after $j_1$ global Grover iterations is given
by
\begin{eqnarray}
G_1^{j_1}|s_1\rangle= \frac{1}{\cos^2\theta_1} \left(
\begin{array}{c}\cos\theta_1
\left( s_g m + c_g\cos\theta_1
\sin\theta_1\right) \\
\cos\theta_2\sin\gamma\left(c_g m-s_g\cos\theta_1 \sin\theta_1
\right)\\
\cos\gamma\left(c_g m-s_g \cos\theta_1 \sin\theta_1 \right)
\end{array}
\right) \!, \label{g1j1}
\end{eqnarray}
where $c_g\equiv\cos(2j_1\theta_1)$, $s_g\equiv\sin(2j_1\theta_1)$
and $m\equiv\cos^2\theta_2\sin^2\gamma+\cos^2\gamma$. For a more
compact presentation of $G_1^{j_1}|s_1\rangle$, let $k\equiv s_g m +
c_g\cos\theta_1 \sin\theta_1$ and $l\equiv c_g m-s_g \cos\theta_1
\sin\theta_1$. Then $G_1^{j_1}|s_1\rangle$ is compactly represented
as follows.
\begin{eqnarray}
G_1^{j_1}|s_1\rangle= \frac{1}{\cos^2\theta_1} \left(
\begin{array}{c}
\cos\theta_1 \left( k \right) \\
\cos\theta_2\sin\gamma\left(l \right)\\
\cos\gamma\left(l\right)
\end{array} \right) \!.
\label{compactg1j1}
\end{eqnarray}
The next intermediate state after $j_1$ global and $j_2$ local
Grover iterations is given by
\begin{eqnarray}
G_2^{j_2}G_1^{j_1}|s_1\rangle  \nonumber \\
=\frac{1}{\cos^2\theta_1} \left(
\begin{array}{c}
c_l\cos\theta_1\left(k\right)+ s_l\cos\theta_2\sin\gamma\left(l\right)  \\
-s_l\cos\theta_1\left(k\right)+ c_l\cos\theta_2\sin\gamma\left(l\right) \\
\cos\gamma\left(l\right)
\end{array} \right)
= \left(\begin{array}{c}a\\ b\\ c\end{array}\right)  \!, \label{14}
\end{eqnarray}
where $c_l\equiv \cos(2j_2\theta_2)$ and $s_l\equiv
\sin(2j_2\theta_2)$.

The final global Grover operator iteration, $-I_sI_t$, is modified
with two phases as in the exact Grover search
\cite{2000quant.ph..5055B}
\begin{eqnarray}
G_1^{\rm final}&\equiv& -\bigl[\hat{I}-(\hat{I}-e^{2i\theta})|s_1
\rangle\langle s_1|\bigr]\times
\bigl[\hat{I}-(\hat{I}-e^{i(\phi-\theta)}) |M\rangle\langle
M|\bigr]\;.
\end{eqnarray}

Translating this into the three basis states supporting the entire
computation we obtain
\begin{eqnarray}
G_1^{\rm final} \nonumber \\
= \left(
    \begin{array}{ccc}
        -e^{i(\phi-\theta)}[1-p\sin^2\gamma \sin^2\theta_2]
        & p\sin^2\gamma\sin\theta_2\cos\theta_2
        & pf\\
        e^{i(\phi-\theta)}p\sin^2\gamma
         \sin\theta_2 \cos\theta_2
        & p\sin^2\gamma \cos^2\theta_2-1
        & pg\\
        e^{i(\phi-\theta)}pf
        & pg
        & p\cos^2\gamma-1\\
    \end{array}
\right),
\end{eqnarray}
where $p\equiv 1-e^{2i\theta}$, $f\equiv \sin\gamma \sin\theta_2
\cos\gamma$, and $g\equiv \sin\gamma \cos\gamma \cos\theta_2$.

Finally then, our aim of a sure success partial search will be
achieved if we can find two phases, $\theta$ and $\phi$, for the
above final global Grover operator which satisfies the condition
\begin{equation}
|\langle x|G_1^{\rm final}G_2^{j_2} G_1^{j_1} |s_1\rangle| = 0 \;.
\end{equation}

The relevant phase condition then reduces to
\begin{eqnarray}
a e^{i(\phi-\theta)}pf+bpg +c[p\cos^2\gamma-1]= 0\;,
\label{phaseCond}
\end{eqnarray}
where $a$, $b$ and $c$ are defined in ~(\ref{14}).

The phase condition may then be rewritten as
\begin{equation}
e^{i(\phi-\theta)}px +py+2z = 0\;, \label{phase_cond_2}
\end{equation}
where
\begin{eqnarray}
x\equiv af,\; y \equiv bg + c\cos^2\gamma,\; z \equiv
-\frac{c}{2}\label{xyz} \;.
\end{eqnarray}
The real and imaginary parts of ~(\ref{phase_cond_2}) may be
simplified to give
\begin{eqnarray}
\sin\phi &=&
-\frac{y}{x}\sin\theta  -\frac{z}{x\sin\theta } \nonumber, \\
\cos\phi&=& -\frac{y}{x}\cos\theta \;. \label{cos}
\end{eqnarray}
Finally, combining these two equations together, we may eliminate
$\phi$ to yield
\begin{equation}
\sin^2\theta = \frac{z^2}{x^2-y^2-2yz} \label{sin2} \;,
\end{equation}
which to have a solution must satisfy
\begin{equation}
x^2 \ge (y+z)^2 \label{ineq} \;.
\end{equation}
There will then be a solution for $\phi$ provided the
right-hand-sides of ~(\ref{cos}) are bounded in absolute value by
unity.

\section{Summary}

In the paper we described a quantum algorithm for partial search in
a database with several target items. A database of $N$ items
separated into $K$ blocks of $b$ items each. The partial search
algorithm finds a target block. We found a simple dependence on
number of target blocks and on number of target items in each block.
In the first half of the paper, we studied the algorithm for large
blocks to understand universal features. In the second half of the
paper we explained how to apply the algorithm in a case of
finite-sized blocks for realistic applications.

\section*{Acknowledgments}

We are grateful for discussion to M.Mosca, M. Heiligman and R.Cleve.
The work was funded by NSF Grant DMS-0503712. BSC was partially
supported by IT Scholarship Program supervised by IITA(Institute for
Information Technology Advancement) \& MIC(Ministry of Information
and Communication), and currently by the Post Brain Korea 21, 2006,
MOE(Ministry of Education \& Human Resources Development), Republic
of Korea.

\end{document}